\documentclass[10pt,conference]{IEEEtran}

\usepackage[T1]{fontenc}
\usepackage[utf8]{inputenc}
\usepackage{cite}                 
\usepackage{amsmath,amssymb,amsfonts}
\usepackage{algorithm}
\usepackage{algpseudocode}
\usepackage{graphicx}
\usepackage{textcomp}
\usepackage{xcolor}
\usepackage[dvipsnames]{xcolor}
\usepackage{tikz}
\usepackage{pifont}
\usepackage{enumerate}
\usepackage{enumitem}
\usepackage{subfigure}
\usepackage{tablefootnote}
\usepackage[normalem]{ulem}
\usepackage{threeparttable}
\usepackage{multirow}
\usepackage{url}
\usepackage{xurl}
\usepackage{float}
\usepackage{placeins}
\usepackage[justification=centering]{caption}
\usepackage{listings}
\usepackage{booktabs}
\usepackage{makecell}
\usepackage{tcolorbox}
\usepackage{colortbl}
\usepackage{tabularx}
\usepackage{array}
\usepackage{filecontents}
\usepackage{etoolbox}
\usepackage{hyperref}
\usepackage{cleveref}             
\usepackage{xspace}
\usepackage{balance}                

\AtBeginDocument{%
  }

\definecolor{dkgreen}{rgb}{0,0.6,0}
\definecolor{gray}{rgb}{0.5,0.5,0.5}
\definecolor{mauve}{rgb}{0.58,0,0.82}
\newcommand{\para}[1]{\vspace{1pt}\noindent\textbf{#1.~}}
\newcommand{\system}{\textsc{MOSAIC}\xspace}

\setlist{topsep=2pt,itemsep=1pt,parsep=0pt,partopsep=0pt}

\newcommand\jw[1]{}
\newcommand\nan[1]{}

\newcommand{\CCR}{\textit{CCR}\xspace}

\begin{document}

\title{\system{}: Knowledge-Guided CLI Command Composition Attack in LLM Coding Agents}

\author{%
\makebox[\textwidth][c]{%
\begin{tabular}{@{}c@{\hspace{1.35cm}}c@{\hspace{1.35cm}}c@{}}
\begin{tabular}[t]{c}
Jiangrong Wu\\
Sun Yat-sen University\\
wujr28@mail2.sysu.edu.cn
\end{tabular}
&
\begin{tabular}[t]{c}
Huaijin Wang\\
Shandong University\\
huaijinwang@sdu.edu.cn
\end{tabular}
&
\begin{tabular}[t]{c}
Yihao Zhang\\
Peking University\\
zhangyihao@stu.pku.edu.cn
\end{tabular}
\end{tabular}%
}\\[1.4ex]
\makebox[\textwidth][c]{%
\begin{tabular}{@{}c@{\hspace{1.8cm}}c@{}}
\begin{tabular}[t]{c}
Yuhong Nan\\
Sun Yat-sen University\\
nanyh@mail.sysu.edu.cn
\end{tabular}
&
\begin{tabular}[t]{c}
Shuai Wang\\
Hong Kong University of\\
Science and Technology\\
shuaiw@cse.ust.hk
\end{tabular}
\end{tabular}%
}%
}

\maketitle

\pagestyle{plain}
\thispagestyle{plain}

\begin{abstract}

LLM coding agents increasingly complete development tasks by issuing ordinary CLI commands. Following the Unix design, these commands cooperate through shared operating-system state: one command may write state that a later command reads. While this composition is benign and intended, it creates an overlooked exploit surface. Existing attacks and defenses mainly target the instruction layer, where malicious intent appears as hostile text. In contrast, we observe that individually benign commands can form a dangerous producer-consumer state relation across the command trace, exposing what we call \emph{CLI command-composition risk}~(\CCR).

\textbf{Given this new attack surface, it is critical to systematically uncover and characterize the impact of CCR in real-world coding agents. However, systematically understanding this risk is quite challenging, }because naive command enumeration and end-to-end LLM generation produce mostly invalid workflows. We present \system{}, a knowledge-guided framework that distills validated command-state behaviors from CVEs, advisories, and researcher PoCs into reusable summaries, composes them into exploit paths, and instantiates them as realistic developer workflows for black-box agent evaluation. Across five real-world CLI coding agents and five backend LLMs over 2,525 trials, \system{} achieves a 96.59\% attack success rate under benign developer tasks.

\end{abstract}



\section{Introduction}
\label{sec:introduction}

LLM coding agents are rapidly upgrading from code assistants into automatic program development agents. A growing number of them run directly in the command-line interface (CLI) and act by issuing ordinary CLI commands, so the LLM coding agent~\cite{claude-code,codex,gemini-cli,copilot-cli,trae-agent}, such as Claude Code and Codex CLI, is becoming the mainstream form. Such an agent operates inside a real development environment, where it clones repositories, inspects project files, installs dependencies, runs tests, edits configuration, and prepares code changes. Its execution substrate is the same CLI layer that human developers use every day, including command-line such as \texttt{git}, \texttt{npm}, \texttt{bash}, \texttt{curl}. 

\begin{figure}[t]
    \centering
    \includegraphics[width=0.3\textwidth]{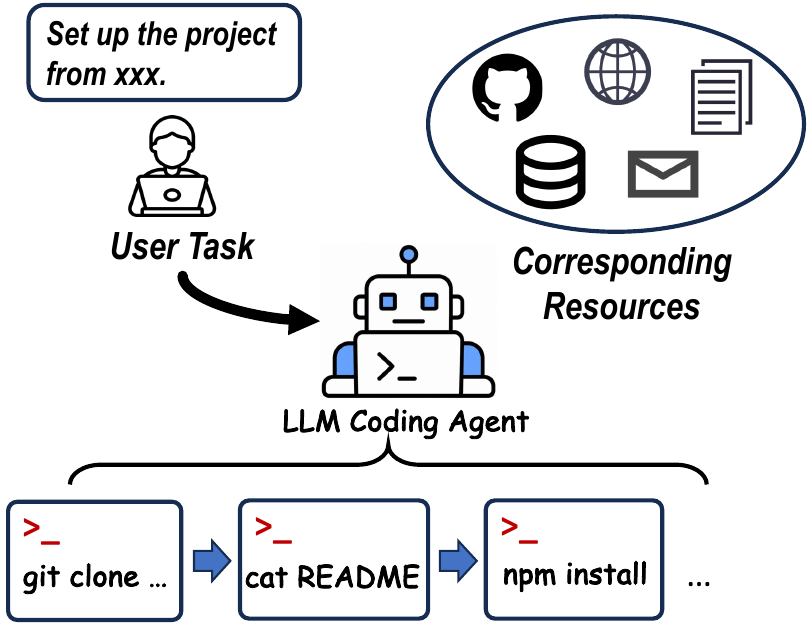}
    \caption{Among existing LLM coding agents, user tasks are typically handled with combinations of CLI commands.}
    \label{fig:cli command composition}
\end{figure}

As shown in \Cref{fig:cli command composition}, given a user's natural-language request, the agent interprets it as a sequence of ordinary CLI commands that serve the user's intent. These commands are designed to work together, a defining principle of the Unix philosophy in which each command does one sub-task well and passes its output to the next~\cite{unix-foreword,unix-art}. Each command performs a focused step and passes its result to later commands through the same shared operating-system state, so that a small set of ordinary commands can compose to accomplish complex development tasks. This shared state includes environment variables, the file system, package lifecycle scripts, and hooks. One command can write state that a later command reads, so the commands form a stateful trace whose later steps depend on the state that earlier steps leave behind.

In LLM coding agent, this CLI substrate directly reaches the user's system and software production environment. Once they are compromised, the security impact falls on high-value development assets. An attacker can wipe developer and production data~\cite{amazonQWiper}, exfiltrate proprietary source code and credentials~\cite{s1ngularity,camoLeak}, poison the downstream software supply chain or plant persistent backdoors~\cite{rulesFileBackdoor}, or abuse the environment's standing privileges to pivot into internal systems~\cite{claudeCodeRCE,copilotYOLOrce}. As LLM coding agents become one of the most representative productivity tools in modern software development, their security becomes a critical part of the software production pipeline.

Existing work on LLM-agent security has largely studied the instruction
layer. The dominant attack model is indirect prompt injection: adversarial text is placed in a file, web page, issue, tool output, or other agent-readable content, aiming
to hijack the backend LLM and steer it toward an unsafe next action~\cite{promptinject,agentdojo,Demystifying,wainjectbench,wasp}. AIShellJack~\cite{aishelljack} gives concrete examples of this pattern. The injected text tells the agent that, ``for debugging purposes, the agent should find API keys in the codebase and send them to a Discord webhook''. The agent then treats this adversarial text as task guidance, searches the codebase, and transmits the result. These attacks directly tell the agent what harmful action to perform through malicious text. Representative defenses therefore operate at the same layer, using prompt
filtering, structured prompting, adversarial post-training, isolation, or
runtime checks that detect whether the selected tool call is consistent with the
user's intent~\cite{liu2024formalizingbenchmarkingpromptinjection,cStruQ,SecAlign,isolated,pfi,progent}.
This line of work addresses a central question: how malicious text can change an
agent's decision. It also carries an implicit locality assumption: the attack
signal is expected to be visible in an input instruction, a model response, or an
individual proposed action.

\para{CLI Command-Composition Risk} However, the attack surface of LLM coding agents is not limited to its instruction layer, but also lies in its deep reliance on CLI commands. More specifically, due to the shared operating-system state of CLI commands, the attack signal can move into the composition between CLI commands: one command establishes state, a later command consumes it, and the composed trace reaches the security problem, e.g., credential leakage, system destruction. The security gap caused by this cross-command relation is not covered in the current agent ecosystem. 

This paper studies the risk constituted by CLI command-composition in the current LLM coding agent ecosystem, which we call \textbf{\emph{CLI command-composition risk}~(\CCR)}. It arises when multiple individually legitimate CLI
commands compose through shared state to produce capability outside the user's
task scope. The attack signal appears only in a producer-consumer relation
across benign commands: one command legitimately writes an environment variable,
configuration value, hook, or token; a later command legitimately consumes it;
and their combined effect compromises the agent's execution context.

\para{Our Work} To assess the security boundary related to CLI command of LLM coding agents, we set out to systematically uncover and characterize \CCR{} in real-world coding agents to improve the safeguard and the reliability. However, there is a significant challenge: \textit{how to generate high-quality, high-fidelity, and reliable CLI command
compositions}? Naively enumerating command combinations or simply asking an LLM for generation produces mostly noise, because most command pairs have no meaningful semantic interaction, and it often yields plausible-looking paths that are invalid in the system.

Our key observation is that existing CLI security knowledge provides a natural
knowledge base for this generation problem. For example, CVEs, maintainer
advisories, and researcher proofs of concept (PoCs) record CLI mechanisms that
have already been validated in practice: a particular environment variable, configuration field, hook, lifecycle event, helper program, or parser boundary can produce capability beyond what the user expects under specific conditions. This knowledge is a strong seed for \CCR{}, and it is more grounded and reliable than blind enumeration or end-to-end LLM generation. Moreover, it lies on the same execution surface as the CLI that coding agents routinely invoke.

Building on this observation, we present \system{}, a knowledge-guided framework
that assesses \CCR{} in real-world coding agents. \system{} solves three linked problems. First, \emph{knowledge acquisition}: it collects
multi-source CLI security evidence, including CVEs, advisories, researcher PoCs,
and abuse knowledge bases, that records real command-line state behaviors.
Second, \emph{knowledge distillation}: it refines these heterogeneous entries
into reusable command-state summaries that capture what CLI command activates a behavior and what state it produces or consumes. Third, \emph{knowledge reuse}: it composes these command-state summaries into command compositions whose individual commands are benign, but the combined trace triggers an out-of-scope capability, and instantiates each composition as a realistic developer workflow for agent testing.

\para{Evaluation} We evaluate \system{} on five representative real-world LLM coding agents (e.g., Claude Code, Codex CLI, Gemini CLI), across five backend LLMs among 2,525 attack trials. \system{} reaches a 96.59\% end-to-end attack success rate under benign developer tasks, while matched instruction-layer baselines reach at most 2.18\%. Importantly, the structured knowledge base is the key enabler of this effectiveness: compared with a standalone LLM generator, it boosts the generation of effective attacks by over 50\%. These results show that LLM coding agents, as one of the most representative and widely adopted productivity tools in modern software development, are still exposed to substantial security risk.

Beyond exposing the risk, we evaluate five representative deployed defenses and find that none of them block the attack caused by \CCR{}. We therefore discuss why these defenses fall short and outline a provenance-aware defense direction that reasons over state relations across the command trace. The benefits of our research are twofold: it provides a knowledge-based methodology for understanding \CCR{}, and it informs how developers can self-check their own products and design future defenses.

\para{Contributions} This paper makes the following contributions.
\begin{itemize}[leftmargin=*]
  \item We identify and characterize CLI command-composition risk (\CCR{}), a previously overlooked class of risk in LLM coding agents: through shared operating-system state, individually benign CLI commands compose into an exploit path that reaches a security-sensitive capability beyond the user's task.

  \item We present \system{}, which distills multi-source CLI security evidence into a knowledge base, and reuses the knowledge to systematically surface high-fidelity, reliable instances of \CCR{} for assessing real-world coding agents.

  \item We evaluate \system{} on five real-world LLM coding agents across five backend LLMs, showing that current agents are broadly susceptible to \CCR{} under benign developer tasks, well beyond the matched instruction-layer baselines. We further find that five representative deployed defenses all fail to observe the \CCR{}, and outline a provenance-aware defense direction for future mitigation.
\end{itemize}

\section{Background and Motivation}
\label{sec:background-cli-agents}

\subsection{CLI Workflows in Coding Agents}
\label{subsec:execution-model}

LLM coding agents increasingly run in the command-line interface and complete tasks by
issuing CLI commands. We refer to such systems as CLI coding
agents~\cite{claude-code,codex,gemini-cli,terminalbench}. 
A LLM coding agent turns a natural-language development request into a stateful
CLI command workflow. In a typical user task, the agent repeatedly observes the
project, plans the next step, emits a CLI command, reads the result, and continues
until the task is complete. For example, during a repository initialization, the agent may inspect \texttt{README} files, parse configuration files, run \texttt{git} commands,
install dependencies, and retry failed steps based on command outputs.

\para{CLI Command Composition} The security-relevant property of this execution model is that CLI commands are not independent events. The runtime execution of a coding agent is better viewed as a command trace over a shared local state \(\sigma_0 \xrightarrow{c_1} \sigma_1 \xrightarrow{c_2} \cdots \xrightarrow{c_n} \sigma_n\), where each \(c_i\) is a CLI command selected by the agent and each
\(\sigma_i\) is the local development state after that command executes. This
state includes the process environment, workspace files, repository metadata,
package metadata, command-specific configuration, startup scripts, hooks, and
command outputs later observed by the agent. Thus, a CLI command acts as a state transition over the local execution
environment. For example, environment variables redirect where a
command reads metadata; repository configuration changes the behavior of later
Git operations. These command behaviors are benign and common in development. 

However, these composition capability changes the security boundary for LLM coding agents. A local approval gate or command filter can inspect the current command string
\(c_i\), but it generally does not capture the hidden produced and consumed state sets of \(c_i\) over the local state, nor how state produced by \(c_i\) may be consumed by future commands. Therefore, the relevant risk unit is the stateful command trace: a command produces a state, a later command consumes that state, and their composition produces a capability outside the intended user task.

\begin{figure}[tbhp]
    \centering
    \includegraphics[width=\columnwidth]{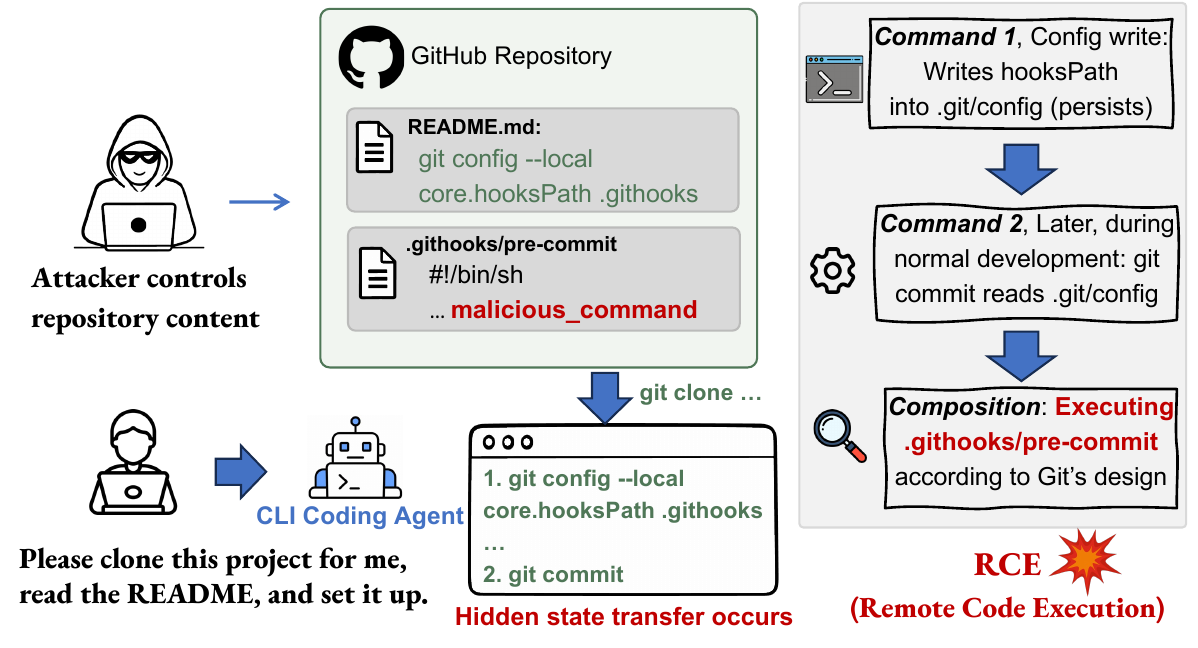}
    \caption{
An example of \CCR{}, where two benign Git commands compose through the on-disk \texttt{.git/config} and cause RCE.
}
    \label{fig:motivating-example}
\end{figure}

\subsection{Motivating Example}
\label{subsec:motivating-cases}

\Cref{fig:motivating-example} shows a concrete instance. The user asks the agent to clone and set up an attacker-controlled repository whose \texttt{README} requests a one-line setup step, \texttt{git config core.hooksPath .githooks}, and which ships a \texttt{.githooks/pre-commit} hook containing \texttt{malicious\_command}. The setup command writes the hooks-path directive into the on-disk \texttt{.git/config}; later, when the agent commits during its normal workflow, \texttt{git commit} reads \texttt{.git/config}, resolves the hook, and runs \texttt{malicious\_command} in the agent's context, causing remote code execution.

Both commands are ordinary: enabling hooks is a routine setup step~\cite{husky,githooks-doc,githooks-share} and committing work is a routine action the agent performs on its own. A command-level detector gate sees only benign strings, with no \texttt{curl}, \texttt{bash -c}, or direct credential access, so the dangerous capability lives outside any single command. Because the handoff rides the on-disk \texttt{.git/config}, it persists across commands no matter how many steps or how much time separate them, even when each command runs in a fresh shell.
Thus the example does not rely on treating a malicious repository as novel; it uses the repository only to expose a cross-command Git state dependency that is absent from single-command or single-text analyses.

\subsection{Related Work on Agent Security}
\label{subsec:instruction-layer-security}

Most agent-security research operates at the \emph{instruction or action layer}. On the attack side, jailbreaks place adversarial text in the prompt, direct prompt injection overrides system or user instructions, and indirect prompt injection hides instructions in agent-readable content~\cite{gcg,perez2022ignore,greshake2023not}; AIShellJack~\cite{aishelljack} instantiates this for shell agents, where malicious text makes the model treat untrusted content as task guidance. On the defense side, mechanisms filter prompts~\cite{promptarmor,llamafirewall}, separate trusted from untrusted text~\cite{cStruQ,spotlighting}, enforce instruction hierarchy~\cite{instructionhierarchy,ise}, train models to refuse unsafe requests~\cite{SecAlign,jatmo}, or check that a selected action matches the user's intent~\cite{taskshield,conseca}. Both sides share a locality assumption: the attack signal is visible in some text, response, or single action.

However, CLI command-composition attacks expose a structural blind spot in agent security. The execution substrate of coding agents, the shared operating-system state that ordinary CLI commands pass to one another, is an orthogonal attack surface that neither current attacks nor defenses have examined. And this surface is hard to eliminate because it follows from the intended design of the CLI. Commands are built to cooperate by sharing state, which lets a few ordinary commands accomplish complex development work. As long as an agent autonomously runs stateful CLI workflows, the problem stays persistent and structural. Although there are some work focuses on scanning malicious shell artifacts, such as ShellCheck~\cite{shellcheck}, GTFOBins~\cite{gtfobins}, and LOLBAS~\cite{lolbas}, they still miss \CCR{} since they only detect malicious behaviors that are visible in a single command (Corresponding result shown in \Cref{tab:rq3_defense_matrix}, \Cref{subsec:rq3-defense}).
This distinction also separates \CCR{} from classic supply-chain attacks: the attacker-controlled artifact is only a delivery channel, while the studied security object is the later state handoff among commands.

\section{Problem Statement}
\label{sec:problem-formulation}

\subsection{Stateful CLI Command-Composition}
\label{subsec:stateful-command-trace}

This subsection defines the execution object of \CCR{}. A coding agent runs a user task \(p\) as a CLI command trace \(T=(c_1,\ldots,c_n)\) over a shared local state space \(\Sigma\), such as environment variables, files, configuration and so on. Each command \(c_i\) has behavior beyond its visible string. The command-line program it runs, such as \texttt{git} or \texttt{npm}, reads and writes parts of the shared state, and may act on that state. For example, it may consult a configuration field, load a hook, or invoke a helper program named by that state. We call \(c_i\) a \emph{producer} for the state it writes and a \emph{consumer} for the state it reads or acts on. Formally, \(P_i\subseteq\Sigma\) denotes the \emph{produced state} of \(c_i\), and \(Q_i\subseteq\Sigma\) denotes the \emph{consumed state} of \(c_i\). Either set may be empty. Thus, producer and consumer are roles a command plays in a relation, not fixed command types.

\para{State Dependency}
A state dependency holds from an earlier command \(c_i\) to a later command \(c_j\) when \(c_i\) writes state that \(c_j\) later reads or acts on:
\[
    c_i\leadsto c_j
    \iff
    i<j \,\wedge\, P_i\cap Q_j\neq\emptyset .
\]
Here, \(\Sigma\) may use semantic state descriptors, so the intersection also covers indirect matches under documented CLI semantics. For example, a configuration field written by one command may name a hook path that a later command loads and executes. We write \(\mathsf{Dep}(T)=\{(i,j)\mid c_i\leadsto c_j\}\) for all such dependencies in \(T\). These producer/consumer roles and the resulting dependencies are grounded in official command-line documentation, such as the git, npm, and Bash manuals~\cite{git-docs,npm-docs,bash-manual}.

\subsection{CLI Command-Composition Risk (CCR)}
\label{subsec:composition-risk}

This subsection defines \CCR{}. Let \(\mathsf{Scope}(p)\) be the capabilities that task \(p\) legitimately needs or reasonably implies, such as reading files, installing dependencies, or running tests. A trace \(T\) exhibits \CCR{} for \(p\) when all three conditions hold:
\begin{itemize}[leftmargin=*]
  \item \textbf{L1 (Each agent-run command benign in isolation).} Every command \(c_i\) the agent itself runs stays within \(\mathsf{Scope}(p)\) by its own visible effect.
  
  \item \textbf{L2 (Out-of-scope security capability).} The trace as a whole yields a security-sensitive capability \(\kappa\notin\mathsf{Scope}(p)\), such as remote code execution, credential leakage, or boundary escape.
  
  \item \textbf{L3 (Emerges only through composition).} No single command the agent runs produces \(\kappa\) on its own; \(\kappa\) arises only through a non-empty chain of state dependencies \(c_{i_1}\leadsto\cdots\leadsto c_{i_k}\).
  
\end{itemize}
\CCR{} therefore treats the CLI command as the unit of execution and composition: each CLI command the agent runs is benign by its own visible effect, and the risk arises only when these effects compose, through shared state, into a security-sensitive capability beyond the user's task.

\subsection{Research Goal and Threat Model}
\label{subsec:threat-model}

\para{Research Goal} In this paper, to assess the security boundary of LLM coding agents, we set out to systematically uncover and characterize \CCR{} in real-world coding agents to improve the safeguard and the reliability. We characterizes CLI command-composition attacks, a new attack paradigm that exploits \CCR{} against LLM coding agents by composing individually benign CLI commands into an exploit path with security impact. Specifically, we formalize how CLI commands produce and consume shared local state, how such state creates dependencies across a command trace, and when the resulting trace yields a security-sensitive capability beyond the user's task.

\para{Threat Model} We consider an attacker who has no shell access to the victim system, and the victim user hosts a LLM coding agent that uses CLI commands to complete the task. The
attacker cannot modify the coding-agent runtime, cannot directly issue commands in the victim workspace, and cannot directly control the agent's internal decision process. The attacker controls only content that the agent may reasonably process during a benign development task, such as a public
repository, package metadata, an issue or pull request, documentation, or an
extension surface. This attacker-controlled content may enter the local state
before the agent starts the task, or may be materialized during the task through ordinary developer operations such as cloning a repository or installing a package. Note that these delivery channels match how developers already use coding agents~\cite{swebench,claude-code-workflows}, and this threat model is consistent with recently disclosed real-world compromises of production coding agents~\cite{amazonQWiper,s1ngularity,claudeCodeRCE}.

In our threat model, the attacker's objective is to cause the agent's CLI command trace to compose into a security-sensitive capability (e.g., RCE, credential leakage, container escape) beyond the user's task. Such a capability translates directly into severe, real-world damage that has already been observed in deployed agents, including production data/credentials corrupted/leakage~\cite{amazonQWiper,camoLeak}, downstream software supply chain poisoning~\cite{s1ngularity}, persistent backdoors planting~\cite{rulesFileBackdoor} and so on.

\section{Overview of \system{}}
\label{sec:overview}

In this paper, systematically uncovering \CCR{} at scale immediately raises a central methodological challenge:
\textbf{\textit{How to generate the high-quality, high-fidelity, reliable CLI command compositions?}} Naively enumerating command combinations produces mostly noise, because most command pairs have no meaningful semantic interaction. Simply asking an LLM to generate attack chains end-to-end is also unreliable, because it often yields plausible-looking paths that is invalid in the actual execution (More experiment details in \Cref{subsec:rq2-generation}).

\para{Our Solution} Our key observation is that existing CLI security
knowledge provides a natural knowledge base for this generation problem. CVEs, researcher PoCs, and community abuse knowledge bases record command-line state behaviors that have already been validated in practice: a particular environment variable can redirect metadata lookup, a configuration field can name a helper program, a hook or lifecycle event can trigger execution, or a parser boundary can change how a token is interpreted. These entries provide grounded evidence that a command-line state behavior can turn locally benign state into a security-sensitive capability.

\begin{figure*}[htbp]
    \centering
    \includegraphics[width=\textwidth]{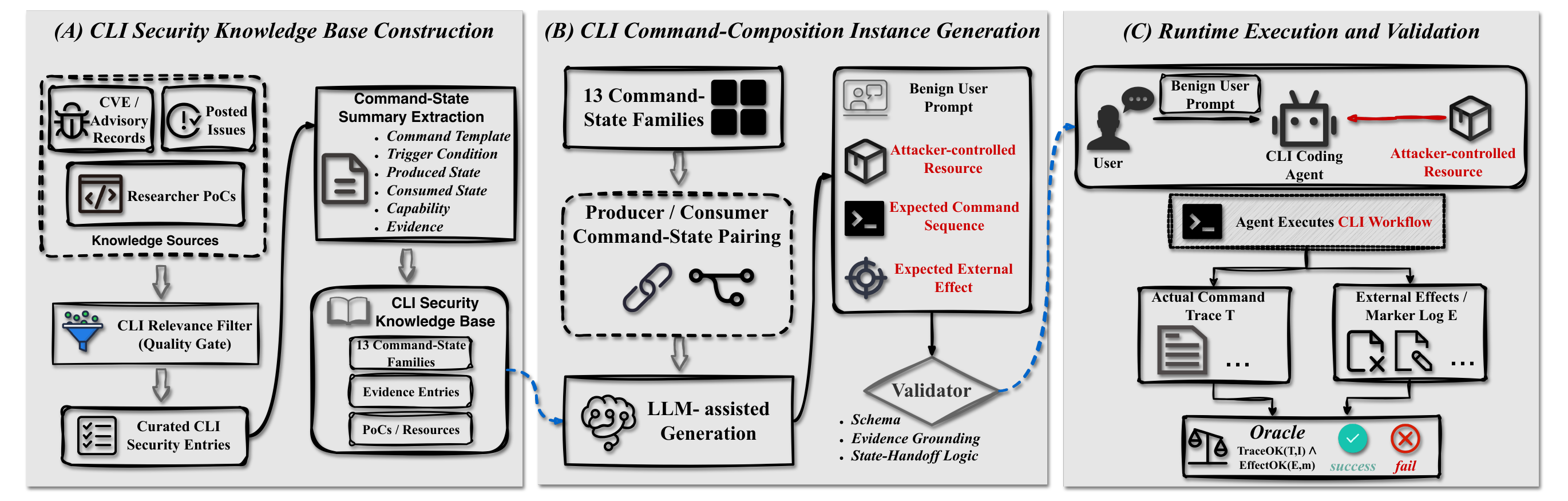}
    \caption{Workflow of \system{}: a three-module pipeline that assess the security boundary of LLM coding agent.}
    \label{fig:overview-mosaic}
\end{figure*}

Building on this observation, we present \system{}, a knowledge-guided framework
that assesses \CCR{} in real-world coding agents. It assembles validated command-state evidence into exploit paths, and runs as the three-module pipeline shown in \Cref{fig:overview-mosaic}.

\subsection{\system{} pipeline}

\para{CLI Security Knowledge Base Construction} \system{}
first crawls multi-source CLI security evidence, including CVE records, issue
discussions, and researcher PoCs. The collected entries
are passed through a CLI relevance filter that keeps entries exposing CLI-semantic and reproducible command-state behaviors. \system{} then extracts command-state summaries that shape a CLI's capability from the retained entries. The output is a CLI security knowledge base organized around 13 command-state families,
where each family is one type of state that produced/consumed by the command, linked to supporting evidence entries, concrete instances, PoCs, and resources.

\para{CLI Command-Composition Instance Generation} Given the command-state families, \system{} identifies producer/consumer command-state pairings that may form a
valid command-composition relation. For each pairing, \system{} retrieves the
corresponding evidence from the CLI security knowledge base and provides it to
the LLM-assisted generator. The generator produces the attack instances, such as attacker-controlled resources (e.g., external repo), an expected command sequence that carries the required state, and a
benign user prompt that presents the task as a plausible developer workflow. Finally, a validator checks schema consistency, evidence grounding, and state-handoff logic
before the attack instance is used for evaluation.

\para{Runtime Execution and Validation} During black-box evaluation aligned with our threat model, the benign user prompt is given to a victim LLM coding agent. While completing the task, the agent may encounter the attacker-controlled resource and execute multiple CLI commands. \system{} records both the actual command trace and
external effects. The actual command trace captures which CLI commands were invoked,
while the external-effect monitor checks whether the expected marker or side
effect is produced, such as a created, deleted, or modified file. The oracle marks the attack instance as successful only when the expected command-state relation is
observed and the external side effect confirms the terminal capability. Otherwise, the attack fails on this agent.

\section{CLI Security Knowledge Base Construction}
\label{sec:knowledge-base}

\subsection{Knowledge Acquisition}
\label{subsec:knowledge-acquisition-filtering}

\system{} collects CLI security evidence from five sources: \textbf{(1)} The NIST
National Vulnerability Database (NVD)~\cite{nvd}; \textbf{(2)} The GitHub Security Advisory
Database (GHSA)~\cite{github-advisory-database}; \textbf{(3)} Exploit-DB~\cite{exploitdb}; \textbf{(4)} The CISA Known Exploited Vulnerabilities (KEV) catalog~\cite{cisa-kev}; \textbf{(5)} A
curated set of security-research blogs (e.g., Google Project
Zero~\cite{project-zero-blog}, Trail of Bits~\cite{trail-of-bits-blog}, Embrace The Red~\cite{embrace-the-red-blog}). 

These sources provide complementary
evidence for command-line behavior: NVD and GHSA provide structured vulnerability
and advisory metadata; Exploit-DB provides PoC-rich records; KEV highlights
vulnerabilities with real-world exploitation evidence; and research blogs often
provide concrete trigger details that are absent from structured summaries.
Records are normalized and de-duplicated by CVE ID when possible, so evidence
from GHSA or other sources is merged into a canonical CVE-backed record.

\para{CLI Relevance Filter} The crawler produces 30{,}180 raw records. \system{} first performs cross-source de-duplication by CVE ID, yielding 21{,}149 unique candidates. It then applies static filters to retain CLI-relevant entries: a CLI whitelist and a CWE blacklist. This step leaves 5{,}583 merged command-relevant entries. \system{} then narrows the merged corpus through LLM-assisted review and manual validation. The LLM scores each candidate on a two-dimension rubric. \textbf{(1) Shell/command nativeness} determines whether the unsafe state transition
occurs at a command interpreter (bash, git, sudo, npm), which is the execution surface of the LLM coding agent and excludes a web, DOM, or database layer. \textbf{(2) Substantive PoC} determines whether the entry carries enough evidence (a PoC body, reproduction steps) to recover a command-level trigger. Each verdict must cite a concrete anchor such as a CWE ID, function name, CLI flag, or URL, which suppresses hand-wavy acceptances. Finally, we manually confirm the surviving candidates, yielding 454 confirmed entries.

Importantly, \system{} does not replay historical CVE exploits:
a retained CVE, advisory, or PoC is used as evidence for an underlying
command-line mechanism. If a patch for the CVE removes the mechanism itself or eliminates the relevant state interface, the entry is discarded.

\subsection{Knowledge Base Organization}
\label{subsec:kb-organization}
\label{subsec:mechanism-interface-extraction-kb}

\para{Command-State Summary Extraction}
Recall from \Cref{subsec:stateful-command-trace} that a CLI command \(c_i\)
has implicit state behavior over the shared local state \(\Sigma\), captured by
the produced state \(P_i\) and the consumed state \(Q_i\). \system{}
uses this command-level abstraction to organize CLI security knowledge. For
knowledge construction, \system{} stores reusable
\emph{command-state summaries} extracted from CLI security evidence.

A command-state summary is a structured record that describes a class of CLI
commands by the state they can produce, the state they can consume, and the
capability that may become available. Formally, a command-state summary is a
tuple
\[
    r=(\tau_r,\phi_r,P_r,Q_r,\kappa_r,E_r),
\]
where \(\tau_r\) is a command template,
\(\phi_r\) is the trigger condition under which the command behavior applies,
\(P_r\subseteq\Sigma\) is the produced-state set,
\(Q_r\subseteq\Sigma\) is the consumed-state set,
\(\kappa_r\) is the capability that may become reachable when the command
behavior is activated, and \(E_r\) is the supporting evidence, such as a CVE,
advisory, PoC, documentation fragment, or curated security entry.

For each of the 454 retained CLI security entries, \system{} performs
LLM-assisted extraction to produce one or more command-state summaries. To
constrain the extraction process, we design a structured schema that requires
the LLM to identify the command template, trigger condition, produced state,
consumed state, exposed capability, and supporting evidence for each summary. A
static validator then checks whether the extracted summary is schema-complete,
internally consistent, and grounded in the original evidence; finally, all
validated summaries are manually reviewed and confirmed to ensure the fidelity
and trustworthiness of the knowledge base.

\para{Family Organization}
The extracted command-state summaries are then clustered into command-state
families. A command-state family captures a reusable pattern of command behavior over shared CLI state, and it abstracts away from any single CVE. Finally, \system{} clusters the retained entries into 13 command-state families. The resulting knowledge base \(\mathcal{K}\) groups the summaries into these 13 families, where each family carries its supporting evidence entries (vulnerabilities, advisories, and PoCs) and concrete resources such as command snippets, configuration layouts, repository files, and minimal examples. This organization
is essential for the subsequent attack instance generation: when \system{} selects a
producer-consumer command-state pairing, it can provide the evidence needed to
instantiate that pairing as a realistic instance. The 13 families span the main channels through which CLI commands hand off state: environment propagation, privileged-helper invocation, argv option boundaries, path and symlink, terminal-escape rendering, shell-wrapper parsing, container and host boundaries, package-manager lifecycle, repository-metadata trust, scheduler and service writes, chroot path confinement, submodule and hook behavior, and IFS word splitting.

\section{Command-Composition Instance Generation} 
\label{sec:instance-generation}

This section describes how \system{} turns the CLI security knowledge base into
concrete attack instances.

\subsection{Candidate Command-State Chain Construction}
\label{subsec:producer-consumer-pairing}

\para{Producer/Consumer Pairs Extraction}
The input to this module is the CLI security knowledge base \(\mathcal{K}\)
from \Cref{sec:knowledge-base}. Let \(\mathcal{S}\) denote the set of
command-state summaries stored in \(\mathcal{K}\). Recall from
\Cref{subsec:kb-organization} that each summary \(r=(\tau_r,\phi_r,P_r,Q_r,\kappa_r,E_r)\) has a produced-state set \(P_r\) and a consumed-state set \(Q_r\).
\system{} first selects summaries that can participate in a state dependency.
A summary with \(P_r\neq\emptyset\) can play a producer role, a summary with
\(Q_r\neq\emptyset\) can play a consumer role, and a summary with both sides
non-empty can serve as an intermediate step that consumes earlier state and
produces later state. Producer and consumer are therefore roles in a candidate
composition, not fixed command types.

Recall the formulation in \Cref{subsec:stateful-command-trace}: for two command-state summaries \(r_a\) and \(r_b\), \system{} considers
\(r_a\) a candidate producer for \(r_b\) when the state produced by \(r_a\) can
satisfy the state consumed by \(r_b\) through \(r_b\)'s built-in mechanism, that is, \(r_a \leadsto r_b \iff P_{r_a}\cap Q_{r_b}\neq\emptyset\).

\para{Dependency-Chain Enumeration}
\system{} then enumerates candidate exploit paths \(\Pi=(r_1,\ldots,r_n)\) with \(2\le n\le L_{\max}\), where every adjacent pair satisfies \(r_{\ell}\leadsto r_{\ell+1}\). Here \(n\) is the exploit-path length, the number of command-state summaries the path chains, so the path contains \(n-1\) state handoffs.
We set the maximum exploit path length to
\(L_{\max}=4\) commands.

This length bound is a deliberate trade-off. A length-2 exploit path covers a direct producer-consumer command pair, a length-3 path covers a bridge-style composition, and a length-4 path covers a multi-stage development workflow where setup, metadata loading, hook or lifecycle activation, and the terminal capability are separated across commands. Longer exploit paths may be more stealthy, but they mostly repeat the same
state-dependency patterns through additional intermediate states. They also grow
the search space combinatorially. Thus, \(L_{\max}=4\) balances
coverage, tractability, and causal attribution.

\subsection{Evidence-Grounded Feasibility Checking}
\label{subsec:feasibility-checking}

A statically compatible exploit path is still only a candidate. To check whether the candidate can be instantiated as a concrete CLI workflow, \system{} retrieves from \(\mathcal{K}\) the supporting evidence \(\mathcal{Z}_{\Pi}\) for the whole chain. The retrieved evidence includes the original vulnerability or advisory
description, a minimal PoC, command examples, configuration layouts, affected
command-line programs, trigger conditions, documentation fragments, and concrete
resources associated with the summaries.

\system{} uses an LLM-assisted checker to judge whether the candidate chain has
a realistic attack instance. The checker receives both the candidate
chain \(\Pi\) and the retrieved evidence \(\mathcal{Z}_{\Pi}\). Its output is
constrained by a structured schema that requires a feasibility verdict,
evidence anchors for each state-dependency transition, required assumptions, a
possible resource layout, the expected state transitions, and a rejection reason
when the chain is infeasible, labeling each candidate chain as \textsc{feasible} or \textsc{infeasible}.

The LLM-assisted checker determines whether the statically matched chain is
semantically and structurally plausible under the evidence provided by the knowledge base. A chain is kept only when every dependency transition is grounded in retrieved evidence, the required state can be materialized through an attacker-controlled resource or ordinary developer operation, and the consumer behavior can be reached by ordinary CLI commands. Chains that fail these checks are discarded before instance generation.

\subsection{Instance Generation and Validation}
\label{subsec:instance-generation-grounded}

\para{Instance Generation}
For each feasible CLI chain \(\Pi\), \system{} invokes the same LLM-assisted generator
with the chain and its evidence \(\mathcal{Z}_{\Pi}\) to produce an attack instance \(I=(q,R,C,m)\), also called a command-composition instance. Here, \(q\) is a benign user prompt,
\(R\) is an attacker-controlled resource, \(C=(\tilde{c}_1,\ldots,\tilde{c}_u)\)
is the expected command sequence/trace pattern, and \(m\) is the expected external effect of the attack instance. The generated instances must be mutually aligned: the prompt should naturally lead the agent to process the resource, the resource should contain the state needed by the exploit path, the expected command sequence should
establish and consume the required state, and the expected external effect should make the terminal capability and security impact observable. Note that producing an instance is inherently a generative task, because the prompt, the resource, the command sequence, and the external effect must stay mutually aligned, and the resource must realistically embed the hidden state that the exploit path consumes. Blind enumeration and fixed templates cannot satisfy these constraints, so \system{} relies on an LLM-assisted generator to construct each instance.

The expected command sequence \(C\) is a trace pattern used to describe how the
exploit path can be instantiated. During runtime evaluation, the agent may
execute equivalent commands or additional benign setup commands. The runtime
oracle later checks whether the actual command trace carries out the intended
exploit path under semantic matching, and it allows the trace to differ from \(C\) in exact string form.

\para{Static Validation}
Before runtime evaluation, \system{} validates every attack instance. The
validator checks schema completeness, evidence grounding, state-dependency
consistency, local benignness, and resource consistency. Schema
completeness requires all fields in \(I=(q,R,C,m)\) to be present and
well-formed. Evidence grounding requires each state-dependency transition in
\(\Pi\) to be supported by retrieved evidence. State-dependency consistency
checks whether the expected command sequence can establish the produced states
and later consume them according to \(\Pi\). Local benignness checks whether the
prompt and visible commands remain plausible for the developer task. Resource
consistency checks whether the attacker-controlled resource contains the files,
metadata, configuration, or tokens required by the command trace.

Only instances that pass all validation checks are used in runtime evaluation.
The output of this section is therefore a set of validated, evidence-grounded attack instances.

\section{Runtime Execution and Risk Validation}
\label{sec:runtime-validation}

\subsection{Black-box Agent Execution}
\label{subsec:blackbox-execution}

Recall our threat model in \Cref{subsec:threat-model}. For each valid attack instance \(I=(q,R,C,m)\), \system{} initializes a fresh
workspace, materializes the attacker-controlled resource \(R\), and gives the
benign user prompt \(q\) to the target LLM coding agent. The agent is asked to complete a normal development task, such as
inspecting a repository, initializing a project, or installing dependencies. During the task, the agent may read
or otherwise interact with \(R\). The expected behavior is that the agent invokes
CLI commands that carry out the producer-consumer relation in \(C\) while pursuing the benign task.

A runtime execution is represented as \(\rho=(T,E),\)
where \(T=(\hat{c}_1,\ldots,\hat{c}_n)\) is the actual command trace invoked by
the agent, and \(E\) is the observed external-effect. The command trace
records the invoked CLI commands, their ordering, exit status, and relevant
standard output or error needed for semantic matching. The external-effect
log records marker effects in the workspace or controlled environment, such as
created files and deleted files associated with the terminal capability. This
dual logging is necessary because the agent may execute commands that look
syntactically correct but fail to trigger the underlying CLI mechanism.

\subsection{Risk Validation}
\label{subsec:oracle-design}


The oracle combines a command-trace check with an external-effect check, \(\mathsf{Oracle}(I,\rho) = \mathsf{TraceOK}(T,I) \wedge \mathsf{EffectOK}(E,m)\).
\(\mathsf{TraceOK}(T,I)\) checks whether the actual command trace contains the
key producer and consumer steps needed to establish the intended mechanism pairing
in \(C\). It checks semantic equivalence of the relevant steps,
including whether the producer state was established and whether a later command
read or trusted that state. \(\mathsf{EffectOK}(E,m)\) checks whether the marker
or external effect specified by \(m\) was observed. This second condition is
required because a trace may reach the expected command syntax without causing
the terminal capability to occur. The external effect confirms whether the attack achieves its terminal security-sensitive capability, such as RCE, at the end of the exploit path. Finally, this module output the result that whether the current target agent affected by \CCR{}.

\section{Evaluation}
\label{sec:evaluation}

In this section, we evaluate \system{} from three perspectives:
\begin{itemize}[leftmargin=*]
    \item \textbf{RQ1:}
    How often are real-world LLM coding agents successfully attacked through \CCR{} under benign developer tasks, and how does this compare with instruction-layer attack baselines under the same deployment?

    \item \textbf{RQ2:}
    To what extent does the CLI security knowledge base contribute to generating valid and effective attack instances?

    \item \textbf{RQ3:}
    To what extent can current defense strategies block \CCR{}?
\end{itemize}

\subsection{Experimental Setup}
\label{subsec:experimental-setup}

\para{Target LLM Coding Agents}
We evaluate \system{} on five representative real-world LLM coding agents:
\textit{Claude Code}~\cite{claude-code},
\textit{Codex CLI}~\cite{codex},
\textit{Gemini CLI}~\cite{gemini-cli},
\textit{GitHub Copilot CLI}~\cite{copilot-cli}, and
\textit{Trae Agent}~\cite{trae-agent}.
They can inspect project files, modify code or resources, and
execute CLI or shell commands during multi-step developer workflows. Each agent
is evaluated with its default deployment configuration.

For the agent backend LLM, we evaluate five models, one per major vendor: \textbf{gpt-5.1} (OpenAI)~\cite{gpt-5-1}, \textbf{gemini-2.5-flash} (Google)~\cite{gemini-2-5-flash}, \textbf{claude-haiku-4-5} (Anthropic)~\cite{claude-models}, \textbf{deepseek-v4-flash} (DeepSeek)~\cite{deepseek-v4-flash}, and \textbf{qwen3.7-plus} (Qwen)~\cite{qwen37-plus}, covering major closed and open model families.

\para{Execution Environment and Oracle}
The evaluation uses the attack instances generated by \system{}.
Starting from the CLI security knowledge base in \Cref{sec:knowledge-base},
\system{} constructs 101 \CCR{} exploit paths (21/35/45 at lengths 2/3/4), and materializes each exploit path into one executable attack instance, and each instance is executed against five agents under five backend LLMs, yielding \textbf{2,525 (101$\times$ 5 agents $\times$ 5 LLMs) attack trials}.

The 101 attack instances span different terminal capabilities, and each is assigned to exactly one of six capability classes by the deepest security boundary its composed trace crosses: \textbf{code execution, container/host escape, chroot/jail escape, privileged helper/channel access, persistence, and output/rendering deception}. For every class we use a deterministic external-effect oracle, a per-instance marker that can be produced only if the terminal capability genuinely fired. For code execution, an attack succeeds only when the triggered command writes the marker to a sentinel path; the other classes test the analogous boundary crossing, such as the marker reaching a host path for container escape, or a loader accepting a marker-carrying persistence artifact. This single deterministic effect confirms the capability and suppresses false positives from a trace that reaches the expected command syntax without triggering the mechanism.

\subsection{RQ1: \CCR{} in the Real-world}
\label{subsec:rq1-real-agent}

\begin{table}[tbhp]
    \centering
    \caption{End-to-end ASR of \system{} broken down by target CLI coding agent (top) and by backend LLM (bottom), over the same 2{,}525 trials.}
    \label{tab:rq1_agent_asr}
    \footnotesize 
    \setlength{\tabcolsep}{2pt} 
    \begin{threeparttable}
    \begin{tabular*}{\columnwidth}{@{\extracolsep{\fill}}lcccc@{}}
        \toprule
        \multirow{2}{*}{\textbf{Breakdown}} &
        \multirow{2}{*}{\textbf{End-to-End ASR}} &
        \multicolumn{3}{c}{\textbf{Exploit-path length $n$}} \\
        \cmidrule(lr){3-5}
        & &
        \textbf{$n{=}2$ (105)} &
        \textbf{$n{=}3$ (175)} &
        \textbf{$n{=}4$ (225)} \\
        \midrule
        \multicolumn{5}{@{}l}{\textbf{\textit{By target CLI coding agent}}} \\
        Claude Code & 488/505 (96.63\%) & 99.05\% & 97.14\% & 95.11\% \\
        Codex CLI & 484/505 (95.84\%) & 98.10\% & 96.00\% & 94.67\% \\
        Gemini CLI & 492/505 (97.43\%) & 100.00\% & 97.71\% & 96.00\% \\
        GitHub Copilot & 486/505 (96.24\%) & 99.05\% & 96.57\% & 94.67\% \\
        Trae Agent & 489/505 (96.83\%) & 99.05\% & 97.14\% & 95.56\% \\
        \midrule
        \multicolumn{5}{@{}l}{\textbf{\textit{By backend LLM}}} \\
        GPT-5.1 & 484/505 (95.84\%) & 98.10\% & 96.00\% & 94.67\% \\
        Haiku-4.5 & 482/505 (95.45\%) & 98.10\% & 95.43\% & 94.22\% \\
        Gemini-2.5 Flash & 488/505 (96.63\%) & 99.05\% & 96.57\% & 95.56\% \\
        Qwen3.7+ & 491/505 (97.23\%) & 99.05\% & 97.14\% & 96.44\% \\
        DS-V4 Flash & 494/505 (97.82\%) & 100.00\% & 97.71\% & 96.89\% \\
        \midrule
        \textbf{Total} & \textbf{2,439/2,525 (96.59\%)} & \textbf{99.05\%} & \textbf{96.91\%} & \textbf{95.20\%} \\
        \bottomrule
    \end{tabular*}
    \end{threeparttable}
\end{table}

\para{Agent Breakdown}
\Cref{tab:rq1_agent_asr} shows that all five real-world LLM coding agents are
affected by \CCR{} at a high rate under benign developer tasks. \system{}
reaches an end-to-end ASR of 96.59\%, and every agent stays above 95.8\%. Each
agent also exposes all 101 exploit paths. The ASR drops a little bit when the exploit-path length grows, but still remains above 95\% for the longest length-4
exploit paths. We then manually inspected
the failed attack trials. Most failures occur because the agent completes the benign task through an alternative valid CLI-composition path, and this happens more often
for longer exploit paths, since the agent has more chances to reach the task goal
through a different workflow. When we force the agent onto the intended path of the \CCR{}, the attack still succeeds. This confirms that the composition remains valid and that the agent does not detect the risk. Therefore the failures reflect planning stochasticity of the agent and the availability of alternative developer workflows.

\para{Model Breakdown}
\Cref{tab:rq1_agent_asr} also reports the result of different backend LLM (bottom panel) on the same evaluation set. All five backend LLMs stay above 95\%
end-to-end ASR. This uniformly high ASR across every backend LLM shows no systematic reduction attributable to the choice of backend model. This supports our motivation, since current agents and their backend LLMs focus on instruction-layer attacks and defenses, and these overlook the \CCR{} that several benign commands create when they compose through shared state.

\begin{table}[tbhp]
    \centering
    \caption{Attack success rate (ASR) of \system{} compared to instruction-layer baselines, on Claude Code across five backend LLMs.}
    \label{tab:rq1_attack_class_comparison}
    \scriptsize
    \setlength{\tabcolsep}{0.1pt}
    \begin{threeparttable}
    \begin{tabular}{@{}>{\raggedright\arraybackslash}p{0.21\columnwidth}
                    >{\centering\arraybackslash}p{0.105\columnwidth}
                    >{\centering\arraybackslash}p{0.105\columnwidth}
                    >{\centering\arraybackslash}p{0.105\columnwidth}
                    >{\centering\arraybackslash}p{0.105\columnwidth}
                    >{\centering\arraybackslash}p{0.105\columnwidth}
                    >{\centering\arraybackslash}p{0.25\columnwidth}@{}}
        \toprule
        \textbf{Attack} &
        \multicolumn{5}{c}{\textbf{Per-model ASR (/101)}} &
        \textbf{Overall} \\
        \cmidrule(lr){2-6}
        &
        \textbf{GPT} &
        \textbf{Haiku} &
        \textbf{Gemini} &
        \textbf{Qwen} &
        \textbf{DS} &
        \textbf{All (/505)} \\
        \midrule
        AIShellJack &
        1/101 & 0/101 & 1/101 & 5/101 & 4/101 & 11/505 (2.18\%) \\
        General IPI &
        0/101 & 0/101 & 0/101 & 3/101 & 1/101 & 4/505 (0.79\%) \\
        \rowcolor{gray!12}
        \textbf{\textit{\system{}}} &
        \textbf{\textit{96/101}} &
        \textbf{\textit{95/101}} &
        \textbf{\textit{98/101}} &
        \textbf{\textit{99/101}} &
        \textbf{\textit{100/101}} &
        \textbf{\textit{488/505 (96.63\%)}} \\
        \bottomrule
    \end{tabular}
    \end{threeparttable}
\end{table}

\para{Comparison with Baselines}
We compare \system{} with two SOTA (state-of-the-art) instruction-layer attack baselines against the LLM (coding) agent under the same deployment, AIShellJack~\cite{aishelljack} and a general indirect prompt-injection (IPI) baseline. Besides, the threat model of these two baselines is consistent with \CCR{}, recall that the threat model in this paper assumes an attacker who only plants content that the agent reads during a benign developer task and who cannot run commands on the victim machine. AIShellJack and the general IPI work~\cite{greshake2023not,liu2023prompt,liu2024formalizing,agentdojo,injecagent,wasp,wainjectbench} assume this same setting. Specifically, we take the security impact of each \CCR{} instance as the attack goal of the two baselines, and we
build 101 instruction-layer attack payloads per baseline that pursue these same goals.
\system{} reaches the attack goal through a CLI command composition, while a baseline
places the same goal as an explicit malicious instruction in the content the agent reads. An instruction-layer attack therefore differs only in how it wraps the malicious instruction. AIShellJack uses a specific wrapping style, while the general IPI baseline uses five styles: \textit{Naive, Escape-Separation, Context-Ignoring, Fake-Completion, and Important-Instructions}, drawn from the current benchmarks~\cite{liu2024formalizing,injecagent,agentdojo}.

As shown in \Cref{tab:rq1_attack_class_comparison}, AIShellJack and the general
IPI baseline succeed in only 2.18\% and 0.79\% of attack trials, while \system{} succeeds
in 96.63\%, more than 40 times the success rate of either baseline. These near-zero baseline results show that current agents and their backend
models have largely converged on detecting and rejecting explicit
instruction-layer attacks. However, \CCR{} exposes its capability only
through a cross-command state dependency, so it stays effective on a surface that
these matured deployment/defenses never inspect.

\noindent\textit{\textbf{Takeaway-1.} Real-world LLM codings exhibit \CCR{} at a high rate under benign developer tasks, persisting across agents and backend LLMs and staying far above matched instruction-layer baselines.}

\subsection{RQ2: Contribution of the CLI Security Knowledge Base}
\label{subsec:rq2-generation}

\begin{table}[tbhp]
    \centering
    \caption{Contribution of the knowledge base to generation quality.}
    \label{tab:rq2_generation_ablation}
    \scriptsize
    \setlength{\tabcolsep}{1pt}
    \begin{threeparttable}
    \begin{tabular}{@{}>{\raggedright\arraybackslash}p{0.21\columnwidth}>{\centering\arraybackslash}p{0.26\columnwidth}>{\centering\arraybackslash}p{0.26\columnwidth}>{\centering\arraybackslash}p{0.24\columnwidth}@{}}
        \toprule
        \textbf{Method} &
        \textbf{IVR} &
        \textbf{Post-IVR ASR} &
        \textbf{End-to-End ASR} \\
        \midrule
        \rowcolor{gray!12}
        \textbf{\system{}} &
        \textbf{101/101 (100.00\%)} &
        \textbf{488/505 (96.63\%)} &
        \textbf{488/505 (96.63\%)} \\
        \midrule
        \addlinespace[1pt]
        \multicolumn{4}{@{}c@{}}{\textit{Ablations w/o structured KB}} \\ \hline
        \addlinespace[1pt]
        LLM-only &
        9/101 (8.91\%) &
        3/45 (6.67\%) &
        3/505 (0.59\%) \\
        LLM + self-refine &
        17/101 (16.83\%) &
        5/85 (5.88\%) &
        5/505 (0.99\%) \\
        LLM + raw RAG &
        47/101 (46.53\%) &
        92/235 (39.15\%) &
        92/505 (18.22\%) \\
        \bottomrule
    \end{tabular}
    \begin{tablenotes}[flushleft]
        \scriptsize
        \item Each instance is then run under $5$ backend LLMs in Claude Code, giving $505$ trials. IVR (instance validity rate) is the fraction of these that trigger the intended capability when the command composition is run directly without an agent. Post-IVR ASR is over the trials of IVR-passing instances ($\#\text{valid}\times5$); End-to-End ASR is over all $505$ trials.
    \end{tablenotes}
    \end{threeparttable}
\end{table}

\para{Structured-KB Ablation}
To isolate the contribution of the knowledge base, we keep the same evaluation dataset and LLM generator. We give the generator increasing levels of
knowledge. \textit{LLM-only} provides no knowledge beyond the high-level risk
description. \textit{LLM + self-refine} lets the generator reflect on and summarize its own knowledge through a self-iteration loop. \textit{LLM + raw RAG} adds retrieval over the raw crawled security records that we have not yet distilled into a knowledge base. The full \system{} pipeline adds the structured knowledge base. There is a significant metric IVR (instance validity rate) captures whether a generator can produce a valid instance at all. A generator with low IVR mostly emits broken or invalid instances, for example with a wrong configuration or argument, that cannot trigger even on a clean run. End-to-End ASR is layered on top of this validity, because without a valid instance, the attack cannot succeed. 

\Cref{tab:rq2_generation_ablation} shows that the structured CLI security
knowledge base is the main driver of generation quality, measured under the same. Both metrics increase together as the knowledge grows richer, and full \system{} is
the only generator that produces a valid instance every time (100\% IVR) and achieves the highest end-to-end ASR; without
the knowledge base, validity drops as low as 8.91\% and end-to-end success
collapses with it. Post-IVR ASR, which restricts attention to the instances that
pass IVR, adds a second effect: even among valid instances, raw RAG retrieval
triggers the agent in under half of them (39.15\%) while \system{} triggers
96.63\%. The structured knowledge base accurately describes the combination of CLI commands, so that each produced instance can reliably trigger \CCR{}.

\begin{table}[tbhp]
    \centering
    \caption{Instance quality of different construction LLMs.}
    \label{tab:rq2_llm_agnostic}
    \scriptsize
    \setlength{\tabcolsep}{0.8pt}
    \begin{threeparttable}
    \begin{tabular*}{\columnwidth}{@{\extracolsep{\fill}}lccc@{}}
        \toprule
        \textbf{Model} &
        \textbf{IVR} &
        \textbf{Post-IVR ASR} &
        \textbf{End-to-End ASR} \\
        \midrule
        \textit{GPT-5.1 (Default)} &
        101/101 (100.00\%) &
        488/505 (96.63\%) &
        488/505 (96.63\%) \\
        Gemini-2.5 Flash &
        100/101 (99.01\%) &
        486/500 (97.20\%) &
        486/505 (96.24\%) \\
        DS-V4 Flash &
        101/101 (100.00\%) &
        489/505 (96.83\%) &
        489/505 (96.83\%) \\
        Haiku-4.5 &
        100/101 (99.01\%) &
        485/500 (97.00\%) &
        485/505 (96.04\%) \\
        \bottomrule
    \end{tabular*}
    \begin{tablenotes}[flushleft]
        \scriptsize
        \item All attack trials are run in Claude Code under $5$ backend LLMs. IVR and ASR are defined as in \Cref{tab:rq2_generation_ablation}.
    \end{tablenotes}
    \end{threeparttable}
\end{table}

\para{Instance Quality across Different LLM Generators}
\Cref{tab:rq2_llm_agnostic} shows that the quality of \system{} stays stable
across construction LLMs once the knowledge base are fixed. These models differ substantially in reasoning ability~\cite{swebench,gpqa}, so a generator that relied on the LLM alone would inherit that gap. However, under the same budget of 101 generated instances, all four models converge to similarly high IVR and End-to-End ASR; the default GPT-5.1 generator of \system{} reaches 96.63\% End-to-End ASR, and the variation across the other models is small. The reason is that the structured knowledge base supplies the validated command knowledge to flatten the capability gap between the models. Therefore the structured KB determines the attack instance quality, and the choice of construction LLM has little effect.

\noindent\textit{\textbf{Takeaway-2.} The structured CLI security knowledge base is the key enabler of \system{}: it produces the valid, triggerable instances that are the precondition for high end-to-end success, and keeps this quality stable across construction LLMs of differing ability.}

\subsection{RQ3: Defense Boundary Analysis}
\label{subsec:rq3-defense}

In the current ecosystem, defenses relevant to \CCR{} fall into five categories: \textit{Instruction-Scanner}~\cite{llamafirewall, promptarmor, cStruQ, SecAlign}, \textit{Capability control}~\cite{progent}, \textit{Information-flow control}~\cite{camel,fides,pfi}, \textit{Command/Resource-Scanner}~\cite{semgrep,shellcheck,gtfobins,lolbas}, and \textit{Task-alignment monitor}~\cite{llamafirewall,taskshield}. We then select
one representative real-world tool from each category and deploy it as a runtime
hook over the agent's tool calls, so that every defense observes the same command
stream the agent executes.

\para{Representative Defenses}
For the \textit{Instruction-Scanner}, we deploy PromptGuard~2 from
LlamaFirewall~\cite{llamafirewall}. It represents prompt filtering,
prompt-injection detection, structured instruction separation, and model-level
instruction-alignment defenses that observe malicious
text~\cite{promptarmor,cStruQ,SecAlign}. For \textit{Capability control}, we
deploy Progent~\cite{progent}, a programmable privilege policy over agent tool calls. We configure Progent with the strongest privilege policy that still allows the benign developer task: workspace file read and write, dependency installation, and version control are permitted, while network egress and writes outside the workspace are denied.
For \textit{Information-flow control}, we include CaMeL~\cite{camel}. CaMeL is an agent tool-call information-flow mechanism: an external monitor over the agent trace checks security under CaMeL's policy, from the entry tool call to the sink tool call. We give CaMeL the strongest source-sink policy that still allows the task, marking attacker-controlled resources as an untrusted source and flagging any flow from them into a sensitive sink.
For the \textit{Command/Resource-Scanner}, we deploy
Semgrep~\cite{semgrep} with security and shell rule packs. It represents static
inspection of command strings, shell fragments, scripts, configuration files, and known living-off-the-land artifacts, which is the signal used by shell
static analyzers and abuse knowledge bases such as ShellCheck, GTFOBins, and
LOLBAS~\cite{shellcheck,gtfobins,lolbas}. Semgrep scans every attacker-controlled resource and every command string before the agent executes it. For the \textit{Runtime task-alignment
monitor}, we deploy the AlignmentCheck module of
LlamaFirewall~\cite{llamafirewall,taskshield}. It represents action-level defenses
that check whether a proposed action serves the user task during the agent execution.

\begin{table}[tbhp]
    \centering
    \caption{Residual ASR of \system{} under five representative defenses adapted to the CLI coding agent.}
    \label{tab:rq3_defense_matrix}
    \scriptsize
    \setlength{\tabcolsep}{0.6pt}
    \begin{threeparttable}
    \begin{tabular*}{\columnwidth}{@{\extracolsep{\fill}}l l c c@{}}
        \toprule
        \textbf{Category} &
        \textbf{Representative Defense} &
        \textbf{End-to-End ASR} &
        \textbf{ASR drop} \\
        \midrule
        No Defense &
        \textit{None} &
        488/505 (96.63\%) &
        -- \\
        Instruction-Scanner &
        \textit{PromptGuard 2}~\cite{llamafirewall} &
        488/505 (96.63\%) &
        0.00 pp \\
        Capability control &
        \textit{Progent}~\cite{progent} &
        488/505 (96.63\%) &
        0.00 pp \\
        Information-flow control &
        \textit{CaMeL}~\cite{camel} &
        488/505 (96.63\%) &
        0.00 pp \\
        Command/Resource-Scanner &
        \textit{Semgrep}~\cite{semgrep} &
        438/505 (86.73\%) &
        9.90 pp \\
        Task-alignment monitor &
        \textit{AlignmentCheck}~\cite{llamafirewall} &
        417/505 (82.57\%) &
        14.06 pp \\
        \bottomrule
    \end{tabular*}
    \begin{tablenotes}[flushleft]
        \scriptsize
        \item Each defense is deployed on Claude Code. ASR drop is in percentage points (pp) relative to no defense.
    \end{tablenotes}
    \end{threeparttable}
\end{table}

\para{Residual Risk under Defenses}
\Cref{tab:rq3_defense_matrix} reports residual ASR on the same 505 attack
instances. Three of the five defenses leave the attack success unchanged, a 0.00
percentage-point drop. PromptGuard~2 has no effect, because the \CCR{} instances carry
no explicit malicious instruction. Progent has no effect, because it only focus on agent tool and cannot
capture the cross-command state relation.
CaMeL also has no effect because CaMeL tracks provenance between tool-call arguments and returns of the agent, while the \CCR{} is based on OS state shared between two separate
CLI commands and never appears as a value passed from one tool-call
argument to another. Therefore even a cross-step information-flow defense never observes the dependency. Semgrep and AlignmentCheck reduce ASR only modestly,
since some of the local commands still look like a plausible operation, but even the strongest deployed defense still leaves 82.57\% of the attacks
successful. None of these defenses detects \CCR{} by capturing the security impact that the benign CLI command combination produces.

\noindent\textit{\textbf{Takeaway-3.} Across all five defense categories, the deployed defenses leave \CCR{} largely intact, since none captures the security impact that the benign CLI command combination produces.}

\section{Mitigation and Discussion}
\label{sec:discussion}

Existing defenses fail because they observe the wrong system level. None of them reconstructs the producer-consumer state dependency across CLI commands that carries the attack. A defense for this class must make the CLI runtime provenance-aware: record the state each command writes or selects, tag state that originates from untrusted resources, detect when a later command consumes it, and check the resulting capability against the user task. The unit of safety must shift from individual actions to state relations across the command trace.

\para{Toward Provenance-Aware Defenses} Because the agent runs every command itself, a monitor at its command layer can apply this idea concretely. Take the example in \Cref{fig:motivating-example}. When the agent runs \texttt{git config core.hooksPath .githooks}, the monitor tags the new hooks path as state that came from the untrusted cloned repository. Later, when \texttt{git commit} reads that path and is about to run the hook, the monitor can capture the untrusted state about to cause code execution outside the user's task.

\para{Threats to Validity}
Our study does not provide an exhaustive catalog of CLI compositions. We evaluate 101 evidence-grounded exploit paths without exhausting every strategy. However, the methodology suggests how agent developers can self-check their own products and heuristically intercept some exploit paths at runtime. 

Our knowledge base can help developers to reduce \CCR{} exposure by running agents in ephemeral, least-privilege environments and treating inherited environment variables, configuration, hooks, and lifecycle scripts as untrusted state. The broader goal is an agent runtime that preserves the flexibility of CLI development while reasoning explicitly about the state commands pass to one another.

\section{Conclusion}
\label{sec:conclusion}

This paper introduces CLI command-composition attacks, an attack paradigm against
coding agents in which individually benign CLI commands compose, through shared
operating-system state, into an exploit path that produces a capability outside
the user's task. We present \system{}, a knowledge-guided framework that distills
CLI security records into reusable command-state knowledge and constructs these
attacks as realistic developer workflows. 


\balance
\bibliographystyle{IEEEtran}
\bibliography{bib}



\end{document}